\documentclass[letterpaper,12pt]{article}
\usepackage[margin=1in]{geometry}
\usepackage{verbatim}
\usepackage{amsmath}
\usepackage{amssymb}
\usepackage{graphicx}
\usepackage{amsthm}
\usepackage{subfig}
\usepackage[round,authoryear]{natbib}

\def\epsilon{\varepsilon}

\def\phi{\varphi}

\def\0s{{\bf 0}}
\def\E{{\bf E \,}}

\newtheorem{theorem}{Theorem}[section]
\newtheorem{lemma}[theorem]{Lemma}

\newtheorem{observe}[theorem]{Observation}
\newtheorem{remark1}[theorem]{Remark}

\newenvironment{remark}{\begin{remark1} \rm}{\end{remark1}}

\title{A comparison of the discrete Kolmogorov-Smirnov statistic
       and the Euclidean distance}
\author{Jacob Carruth, Mark Tygert, and Rachel Ward}

\begin{document}

\maketitle

\begin{abstract}
Goodness-of-fit tests gauge whether a given set of observations
is consistent (up to expected random fluctuations)
with arising as independent and identically distributed (i.i.d.)\ draws
from a user-specified probability distribution known as the ``model.''
The standard gauges involve the discrepancy between the model
and the empirical distribution of the observed draws.
Some measures of discrepancy are cumulative; others are not.
The most popular cumulative measure is the Kolmogorov-Smirnov statistic;
when all probability distributions under consideration are discrete,
a natural noncumulative measure is the Euclidean distance
between the model and the empirical distributions.
In the present paper, both mathematical analysis and its illustration
via various data sets indicate that the Kolmogorov-Smirnov statistic tends
to be more powerful than the Euclidean distance
when there is a natural ordering for the values that the draws can take
--- that is, when the data is ordinal ---
whereas the Euclidean distance is more reliable and more easily understood
than the Kolmogorov-Smirnov statistic when there is no natural ordering
(or partial order) --- that is, when the data is nominal.

\medskip
\smallskip

\end{abstract}

\bigskip

\tableofcontents

\section{Introduction}
\label{intro}

Testing goodness-of-fit is one of the foundations of modern statistics,
as elucidated by~\cite{rao}, for example.
The formulation in the discrete setting involves $n$
independent and identically distributed (i.i.d.)\ draws
from a probability distribution over $m$ bins
(``categories,'' ``cells,'' and ``classes'' are common synonyms for ``bins'').
In accordance with the standard conventions,
we will use $p$ to denote the actual (unknown) underlying distribution
of the draws; $p = (p^{(1)}, p^{(2)}, \dots, p^{(m)})$,
with $p^{(1)}$,~$p^{(2)}$, \dots, $p^{(m)}$ being nonnegative and
\begin{equation}
\label{prob}
\sum_{j=1}^m p^{(j)} = 1.
\end{equation}
We will use $p_0$ to denote a user-specified distribution,
usually called the ``model'';
again $p_0 = (p_0^{(1)}, p_0^{(2)}, \dots, p_0^{(m)})$,
with $p_0^{(1)}$,~$p_0^{(2)}$, \dots, $p_0^{(m)}$ being nonnegative and
\begin{equation}
\label{prob0}
\sum_{j=1}^m p_0^{(j)} = 1.
\end{equation}
A goodness-of-fit test produces a value --- the ``P-value'' ---
that gauges the consistency of the observed data with the assumption
that $p = p_0$.
In many formulations, the user-specified model $p_0$ consists
of a family of probability distributions parameterized by $\theta$,
where $\theta$ can be integer-valued, real-valued, complex-valued,
vector-valued, matrix-valued, or any combination of the many possibilities.
In such cases, the P-value gauges the consistency of the observed data
with the assumption that $p = p_0(\hat\theta)$,
where $\hat\theta$ is an estimate
(taken to be the maximum-likelihood estimate throughout the present paper).
We now review the definition of P-values.

P-values are defined via the empirical distribution $\hat{p}$,
where $\hat{p} = (\hat{p}^{(1)}, \hat{p}^{(2)}, \dots, \hat{p}^{(m)})$,
with $\hat{p}^{(j)}$ being the proportion of the $n$ observed draws
that fall in the $j$th bin, that is, $\hat{p}^{(j)}$ is the number of draws
falling in the $j$th bin, divided by $n$.
P-values involve a hypothetical experiment taking $n$ i.i.d.\ draws
from the assumed actual underlying distribution $p = p_0(\hat\theta)$.
We denote by $\hat{P}$ the empirical distribution of the draws
from the hypothetical experiment;
we denote by $\hat\Theta$ a maximum-likelihood estimate of $\theta$
obtained from the hypothetical experiment.
The P-value is then the probability that the discrepancy
between the random variables $\hat{P}$ and $p_0(\hat\Theta)$
is at least as large as the observed discrepancy
between $\hat{p}$ and $p_0(\hat\theta)$,
calculating the probability under the assumption that $p = p_0(\hat\theta)$.

To complete the definition of P-values, we must choose a measure
of discrepancy. In the present paper, we consider
the (discrete) Kolmogorov-Smirnov and Euclidean distances,
\begin{equation}
\label{dKolmogorov-Smirnov}
d_1(a,b) = \max_{1 \le k \le m}
\left| \sum_{j=1}^k a^{(j)} - \sum_{j=1}^k b^{(j)} \right|
\end{equation}
and
\begin{equation}
\label{dEuclidean}
d_2(a,b) = \sqrt{\sum_{j=1}^m (a^{(j)} - b^{(j)})^2},
\end{equation}
respectively.
The P-value for the Kolmogorov-Smirnov statistic is the probability that
$d_1(\hat{P},p_0(\hat\Theta)) \ge d_1(\hat{p},p_0(\hat\theta))$;
the P-value for the Euclidean distance is the probability that
$d_2(\hat{P},p_0(\hat\Theta)) \ge d_2(\hat{p},p_0(\hat\theta))$.
When evaluating the probabilities, we view $\hat{P}$ and $\hat\Theta$
as random variables, constructed with i.i.d.\ draws
from the assumed distribution $p = p_0(\hat\theta)$,
while viewing the observed $\hat{p}$ and $\hat\theta$ as fixed, not random.

If a P-value is very small, then we can be confident
that the given observed draws are inconsistent with the assumed model,
are not i.i.d.,\ or are both inconsistent and not i.i.d.

Needless to say, the Kolmogorov-Smirnov distance defined
in~(\ref{dKolmogorov-Smirnov}) is the maximum absolute difference
between cumulative distribution functions.
The Kolmogorov-Smirnov statistic depends on the ordering of the bins,
unlike the Euclidean distance.

As supported by the investigations below, we recommend
using the Kolmogorov-Smirnov statistic when there is a natural ordering
of the bins, while the Euclidean distance
is more reliable and more easily understood
than the Kolmogorov-Smirnov statistic when there is no natural ordering
(or partial order).
Unlike the Euclidean distance, the Kolmogorov-Smirnov statistic
utilizes the information in a natural ordering of the bins,
when the latter is available.
\cite{horn} gave similar recommendations when comparing the $\chi^2$
and Kolmogorov-Smirnov statistics.
Detailed comparisons between the Euclidean distance
and $\chi^2$ statistics are available in~\cite{perkins-tygert-ward3}.

The Kolmogorov-Smirnov statistic is cumulative;
it accentuates low-frequency differences
between the model and the empirical distribution of the draws,
but tends to average away and otherwise obscure high-frequency differences.
Similar observations have been made by~\cite{pettitt-stephens},
\cite{dagostino-stephens}, \cite{choulakian-lockhart-stephens},
\cite{from}, \cite{best-rayner}, \cite{haschenburger-spinelli},
\cite{steele-chaseling}, \cite{lockhart-spinelli-stephens}, \cite{ampadu},
and~\cite{ampadu-wang-steele}, among others.
Our suggestions appear to be closest to those of~\cite{horn}.

There are many cumulative approaches similar
to the Kolmogorov-Smirnov statistic.
These include the Cram\'er--von-Mises, Watson, Kuiper, and R\'enyi statistics,
as well as their Anderson-Darling variants;
Section~14.3.4 of~\cite{press-teukolsky-vetterling-flannery},
\cite{stephens2}, and~\cite{renyi} review these statistics.
We ourselves are fond of the Kuiper approach.
However, the present paper focuses on the popular Kolmogorov-Smirnov statistic;
the Cram\'er--von-Mises, Watson, and Kuiper variants are very similar.

The remainder of the present paper has the following structure:
Section~\ref{nonatorder} describes how the Euclidean distance
is generally preferable to the Kolmogorov-Smirnov statistic
when there is no natural ordering (or partial order) of the bins.
Section~\ref{natorder} describes how the Kolmogorov-Smirnov statistic
is generally preferable to the Euclidean distance
when there is a natural ordering of the bins.
Section~\ref{data_analysis} illustrates both cases with examples
of data sets and the associated P-values,
computing the P-values via Monte-Carlo simulations
with guaranteed error bounds.
The reader may wish to begin with Section~\ref{data_analysis},
referring back to earlier sections as needed.

\section{The case when the bins do not have a natural order}
\label{nonatorder}

The Euclidean distance is generally preferable
to the Kolmogorov-Smirnov statistic
when there is no natural ordering (or partial order) of the bins.
As discussed by~\cite{perkins-tygert-ward2},
the interaction of parameter estimation and the Euclidean distance
is easy to understand and quantify, at least asymptotically,
in the limit of large numbers of draws.
In contrast, the interaction of parameter estimation
and the Kolmogorov-Smirnov statistic can be very complicated,
though \cite{choulakian-lockhart-stephens}
and~\cite{lockhart-spinelli-stephens}
have pointed out that the interaction is somewhat simpler
with Cram\'er's and von Mises', Watson's, and some of Anderson's and Darling's
very similar statistics.
That said, the Euclidean distance can be more reliable even when there
are no parameters in the model, that is, when the model $p_0$
is a single, fixed, fully specified probability distribution;
the remainder of the present section describes why.

The basis of the analysis is the following lemma, a reformulation
of the fact that the expected maximum absolute deviation from zero
of the standard Brownian bridge is $\sqrt{\pi/2} \cdot \ln(2) \approx .8687$
\citep[see, for example, Section~3 of][]{marsaglia-tsang-wang}.
\begin{lemma}
\label{bridge}
Suppose that $m$ is even and that $D^{(1)}$,~$D^{(2)}$, \dots, $D^{(m)}$
form a randomly ordered list of $m/2$ positive ones and $m/2$ negative ones
(with the ordering drawn uniformly at random).
Then,
\begin{equation}
\E \max_{1 \le k \le m} \left| \sum_{j=1}^k D^{(j)} \right| \Bigg/ \sqrt{m}
\quad \longrightarrow \quad \sqrt{\pi/2} \cdot \ln(2)
\end{equation}
in the limit that $m \to \infty$, where (as usual)
$\E$\,produces the expected value.
\end{lemma}

We denote by $p$ the actual underlying distribution
of the $n$ observed i.i.d.\ draws.
We denote by $p_0$ the model distribution.
We denote by $\hat{P}$ the empirical distribution
of the $n$ draws. These are all probability distributions, that is,
$p^{(j)} \ge 0$, $p_0^{(j)} \ge 0$, and $\hat{P}^{(j)} \ge 0$
for $j = 1$,~$2$, \dots, $m$, and (\ref{prob}) and~(\ref{prob0}) hold.

Suppose that the actual underlying distribution
$p^{(1)}$,~$p^{(2)}$, \dots, $p^{(m)}$ of the draws
is the same as the model distribution
$p_0^{(1)}$,~$p_0^{(2)}$, \dots, $p_0^{(m)}$;
the random variables $\hat{P}^{(1)}$,~$\hat{P}^{(2)}$, \dots, $\hat{P}^{(m)}$
are then the proportions of $n$ i.i.d.\ draws from $p_0$ that fall
in the respective $m$ bins.
The Euclidean distance is
\begin{equation}
U = \sqrt{\sum_{j=1}^m (\hat{P}^{(j)} - p_0^{(j)})^2}.
\end{equation}
The Kolmogorov-Smirnov statistic is
\begin{equation}
V = \max_{1 \le k \le m} \left|\sum_{j=1}^k (\hat{P}^{(j)} - p_0^{(j)})\right|.
\end{equation}
The expected value of the square of the Euclidean distance is
\begin{equation}
\label{expectedx}
\E U^2 = \sum_{j=1}^m \E(\hat{P}^{(j)} - p_0^{(j)})^2
       = \sum_{j=1}^m \frac{p_0^{(j)}}{n}
       = \frac{1}{n}.
\end{equation}
As shown, for example, by~\cite{durbin} using Lemma~\ref{bridge} above,
the expected value of $\sqrt{n}$ times the Kolmogorov-Smirnov statistic is
\begin{equation}
\label{expecteds}
\E V\sqrt{n} \to \sqrt{\pi/2} \cdot \ln(2) \approx .8687
\end{equation}
in the limit that $n \to \infty$ and $\max_{1 \le j \le m} p_0^{(j)} \to 0$.
Comparing~(\ref{expectedx}) and~(\ref{expecteds}), we see that $U$ and $V$
are roughly the same size (inversely proportional to $\sqrt{n}$)
when the actual underlying distribution of the draws
is the same as the model distribution.

However, when the actual underlying distribution of the draws differs
from the model distribution, the Euclidean distance and the Kolmogorov-Smirnov
statistic can be very different.
If the number $n$ of draws is large, then the empirical distribution $\hat{P}$
will be very close to the actual distribution $p$.
Therefore, to study the performance of the goodness-of-fit statistics
as $n \to \infty$ when the actual distribution $p$ differs
from the model distribution $p_0$ (and both are independent of $n$),
we can focus on the difference between $p$ and $p_0$
(rather than the difference between $\hat{P}$ and $p_0$).
We now define and study the difference
\begin{equation}
\label{diffs}
d^{(j)} = p^{(j)} - p_0^{(j)}
\end{equation}
for $j = 1$,~$2$, \dots, $m$.
The Euclidean distance between $p$ and $p_0$ (the root-sum-square difference)
is
\begin{equation}
\label{Euclidean}
u = \sqrt{\sum_{j=1}^m (d^{(j)})^2}.
\end{equation}
The Kolmogorov-Smirnov statistic (the maximum absolute cumulative difference)
is
\begin{equation}
\label{KS}
v = \max_{1 \le k \le m} \left| \sum_{j=1}^k d^{(j)} \right|.
\end{equation}

For simplicity
(and because the following analysis generalizes straightforwardly),
let us consider the illustrative case in which
$|d^{(1)}| = |d^{(2)}| = \dots = |d^{(m)}|$, that is,
\begin{equation}
\label{equal}
|d^{(j)}| = c_m
\end{equation}
for all $j = 1$,~$2$, \dots, $m$, where $c_m$ is a positive real number
($c_m$ must always satisfy $m \cdot c_m \le 2$,
since $m \cdot c_m = \sum_{j=1}^m c_m = \sum_{j=1}^m |d^{(j)}|
\le \sum_{j=1}^m [p^{(j)} + p_0^{(j)}] = 2$).
Combining~(\ref{diffs}), (\ref{prob}), and~(\ref{prob0}) yields that
\begin{equation}
\label{zero}
\sum_{j=1}^m d^{(j)} = 0.
\end{equation}
Together, (\ref{zero}) and~(\ref{equal}) imply that $m$ is even
and that half of $d^{(1)}$,~$d^{(2)}$, \dots, $d^{(m)}$ are equal to $+c_m$,
and the other half are equal to $-c_m$.

Combining~(\ref{equal}) and~(\ref{Euclidean}) yields
that the Euclidean distance is
\begin{equation}
u = \sqrt{m} \cdot c_m.
\end{equation}
The fact that half of $d^{(1)}$,~$d^{(2)}$, \dots, $d^{(m)}$ are equal
to $+c_m$, and the other half are equal to $-c_m$, yields that
the Kolmogorov-Smirnov statistic $v$ defined in~(\ref{KS})
could be as small as $c_m$ or as large as $m \cdot c_m/2$,
depending on the ordering of the signs
in $d^{(1)}$,~$d^{(2)}$, \dots, $d^{(m)}$.
If all orderings are equally likely (which is equivalent to ordering
the bins uniformly at random), then by Lemma~\ref{bridge} the mean value
for $v$ is
$\sqrt{m\pi/2} \cdot \ln(2) \cdot c_m \approx \sqrt{m} \cdot .8687 \cdot c_m$
in the limit that $m$ is large
(this is the expected maximum absolute deviation from zero
of a tied-down random walk with $m$ steps, each of length $c_m$,
that starts and ends at zero; the random walk ends at zero due to~(\ref{zero})).

Thus, in the limit that the number $n$ of draws is large
(and $\max_{1 \le j \le m} p_0^{(j)} \to 0$, while both the model $p_0$
and the alternative distribution $p$ are independent of $n$),
the Euclidean distance and the Kolmogorov-Smirnov statistic
have similar statistical power on average,
if all orderings of the bins are equally likely.
However, the Euclidean distance is the same for any ordering of the bins,
whereas the power of the Kolmogorov-Smirnov statistic depends strongly
on the ordering.
We see, then, that the Euclidean distance is more reliable
than the Kolmogorov-Smirnov statistic
when there is no especially natural ordering for the bins.

\begin{remark}
\label{l1}
It is possible to use an ordering for which the Kolmogorov-Smirnov statistic
attains its greatest value (this corresponds to renumbering the bins such that
the differences $D^{(j)} = \hat{P}^{(j)}-p_0^{(j)}$ satisfy
$D^{(1)} \ge D^{(2)} \ge \dots \ge D^{(m)}$ or
$D^{(1)} \le D^{(2)} \le \dots \le D^{(m)}$).
However, this data-dependent ordering produces a statistic
which is proportional to the $l^1$ distance $\sum_{j=1}^m |D^{(j)}|$
(whereas the Euclidean distance is the $l^2$ distance),
as remarked at the top of page~396 of~\cite{hoeffding}.
The resulting statistic is no longer cumulative.
\end{remark}

\section{The case when the bins have a natural order}
\label{natorder}

The Kolmogorov-Smirnov statistic is often preferable
to the Euclidean distance when there is a natural ordering of the bins.
In fact, the Kolmogorov-Smirnov statistic is always preferable
when the data is very sparse and there is a natural ordering of the bins.
In the limit that the maximum expected number of draws per bin tends to zero,
the Euclidean distance always takes the same value under the null hypothesis,
providing no discriminative power: indeed, when the draws producing
the empirical distribution $\hat{P}$ are taken
from the model distribution $p_0$,
the Euclidean distance is almost surely $1/\sqrt{n}$,
\begin{equation}
\sqrt{\sum_{j=1}^m (\hat{P}^{(j)}-p_0^{(j)})^2} = \frac{1}{\sqrt{n}},
\end{equation}
in the limit that $n \cdot \max_{1 \le j \le m} p_0^{(j)} \to 0$
(the reason is that, in this limit, $\max_{1 \le j \le m} p_0^{(j)} \to 0$
and moreover almost every realization of the experiment
satisfies that, for all $j = 1$,~$2$, \dots,~$m$,
$\hat{P}^{(j)} = 0$ or $\hat{P}^{(j)} = 1/n$, that is,
there is at most one observed draw per bin).
In contrast, the Kolmogorov-Smirnov statistic is nontrivial even in the limit
that the maximum expected number of draws per bin tends to zero --- in fact,
this is exactly the continuum limit
for the original Kolmogorov-Smirnov statistic
involving continuous cumulative distribution functions
(as opposed to the discontinuous cumulative distribution functions arising
from the discrete distributions considered in the present paper).
Furthermore, the Kolmogorov-Smirnov statistic is sensitive to symmetry
(or asymmetry) in a distribution, and can detect other interesting properties
of distributions that depend on the ordering of the bins.

\section{Data analysis}
\label{data_analysis}

This section gives four examples illustrating the performance
of the Kolmogorov-Smirnov statistic and the Euclidean distance
in various circumstances.
The Kolmogorov-Smirnov statistic is more powerful than the Euclidean distance
in the first two examples, for which there are natural orderings of the bins.
The Euclidean distance is more reliable than the Kolmogorov-Smirnov statistic
in the last two examples,
for which any ordering of the bins is necessarily rather arbitrary.
We computed all P-values via Monte-Carlo simulations
with guaranteed error bounds, as in Remark~3.3 of~\cite{perkins-tygert-ward3}.
Remark~3.4 of~\cite{perkins-tygert-ward3} proves that the standard error
of the estimate for a P-value $P$ is $\sqrt{P(1-P)/\ell}$,
where $\ell$ is the number of simulations conducted to calculate the P-value.

\subsection{A test of randomness}

A particular random number generator is supposed to produce an integer
from 1 to $2^{32}$ uniformly at random.
The model distribution for such a generator is
\begin{equation}
\label{simplemod}
p_0^{(j)} = 2^{-32}
\end{equation}
for $j = 1$,~$2$, \dots, $2^{32}$.
We test the (obviously poor) generator which produces the numbers
1, 2, 3, \dots, $n$, in that order, so that the observed distribution
of the generated numbers is
\begin{equation}
\label{baddata}
\hat{p}^{(j)} = \left\{ \begin{array}{rl} 1/n, & j = 1,\ 2,\ \dots,\ n \\
                                            0, & j = n+1,\ n+2,\ \dots,\ 2^{32}
                        \end{array} \right.
\end{equation}
for $j = 1$,~$2$, \dots, $2^{32}$.
For these observations, the P-value for the Euclidean distance is 1
to several digits of precision,
while the P-value for the Kolmogorov-Smirnov statistic is 0
to several digits, at least for $n$ between a hundred and a million.
So, as expected, the Euclidean distance has almost no discriminative power
for such sparse data, whereas the Kolmogorov-Smirnov statistic
easily discerns that the data~(\ref{baddata}) is inconsistent
with the model~(\ref{simplemod}).

\begin{remark}
Like the Euclidean distance, classical goodness-of-fit statistics
such as $\chi^2$, $G^2$ (the log--likelihood-ratio),
and the Freeman-Tukey/Hellinger distance are invariant
to the ordering of the bins,
and also produce P-values that are equal to 1 to several digits of precision,
at least for $n$ between a hundred and a million. 
For definitions and further discussion of the $\chi^2$, $G^2$,
and Freeman-Tukey statistics, see Section~2 of~\cite{perkins-tygert-ward3}.
\end{remark}

\subsection{A test of Poissonity}

A Poisson-distributed random number generator with mean $100$ is supposed
to produce a nonnegative integer according to the model
\begin{equation}
\label{models}
p_0^{(j)} = \frac{100^j}{j! \cdot \exp(100)}
\end{equation}
for $j = 0$,~$1$,~$2$,~$3$, \dots.
We test the (obviously poor) generator which produces
the numbers 100, 101, 102, \dots, 109,
so that the observed distribution of the numbers is
\begin{equation}
\label{observations}
\hat{p}^{(j)}
= \left\{ \begin{array}{rl} 1/10, & j = 100, 101, 102, \dots, 109 \\
                               0, & \hbox{otherwise}
          \end{array} \right.
\end{equation}
for $j = 0$,~$1$,~$2$,~$3$, \dots.
The P-values, each computed via 4,000,000 simulations, are
\begin{itemize}
\item Kolmogorov-Smirnov: .0075
\item Euclidean distance: .998
\item $\chi^2$: .999
\item $G^2$ (the log--likelihood-ratio): .999
\item Freeman-Tukey (the Hellinger distance): .998
\end{itemize}
For definitions and further discussion of the $\chi^2$, $G^2$,
and Freeman-Tukey statistics, see Section~2 of~\cite{perkins-tygert-ward3}.
The Kolmogorov-Smirnov statistic is far more powerful for this example,
in which the bins have a natural ordering
(in this example the bins are the nonnegative integers).

\begin{figure}[p]
\begin{center}
\rotatebox{-90}{\scalebox{.47}{\includegraphics{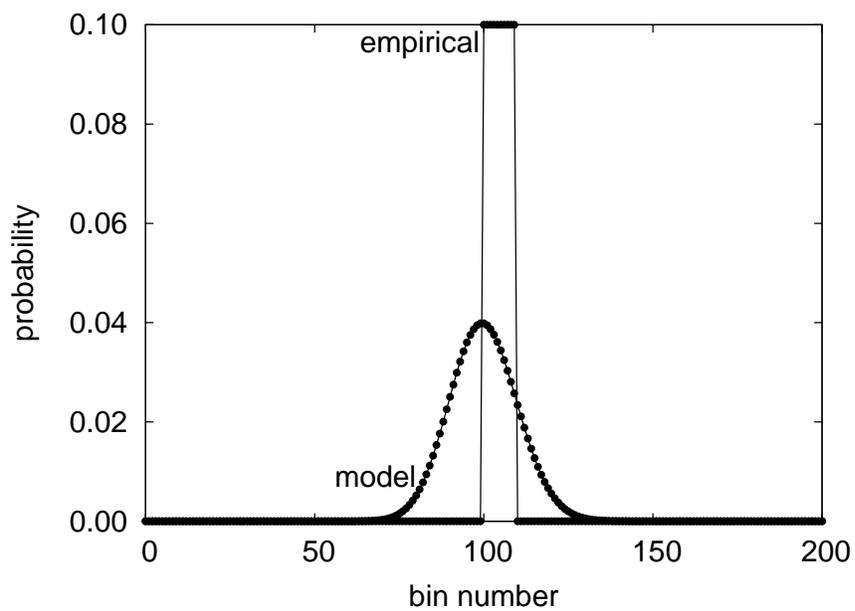}}}
\\\vspace{.1in}
\caption{Proportions associated with the bins for the observations}
\label{observedpmf}
\end{center}
\end{figure}

\begin{figure}
\begin{center}
\rotatebox{-90}{\scalebox{.47}{\includegraphics{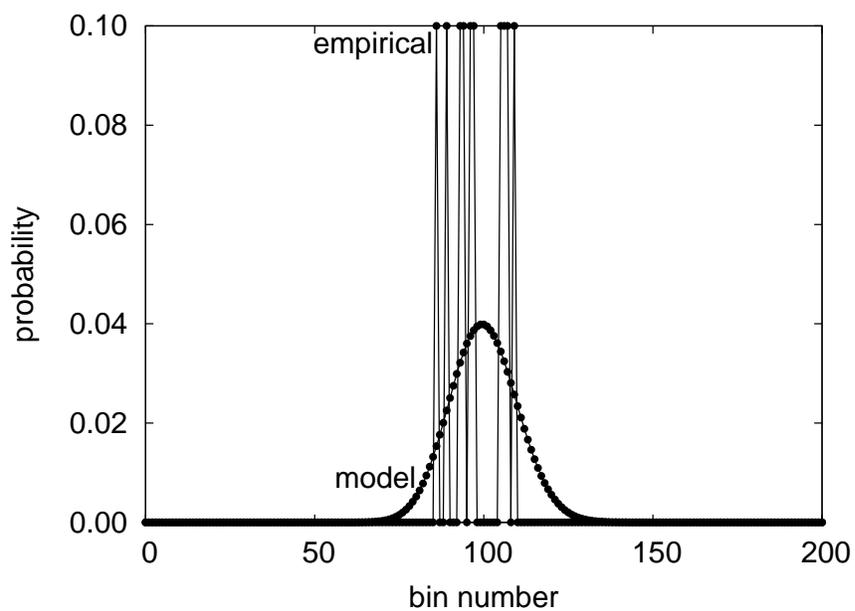}}}
\\\vspace{.1in}
\caption{Proportions associated with the bins for a simulation}
\label{simulatedpmf}
\end{center}
\end{figure}

Figure~\ref{observedpmf} plots the model probabilities
$p_0^{(0)}$, $p_0^{(1)}$, $p_0^{(2)}$, \dots\ defined in~(\ref{models})
along with the observed proportions
$\hat{p}^{(0)}$, $\hat{p}^{(1)}$, $\hat{p}^{(2)}$, \dots\ defined
in~(\ref{observations}).
Figure~\ref{simulatedpmf} plots the model probabilities
$p_0^{(0)}$, $p_0^{(1)}$, $p_0^{(2)}$, \dots\
along with analogues of the proportions
$\hat{p}^{(0)}$, $\hat{p}^{(1)}$, $\hat{p}^{(2)}$, \dots\
for a simulation generating 10 i.i.d.\ draws according to the model.

\begin{figure}
\begin{center}
\rotatebox{-90}{\scalebox{.47}{\includegraphics{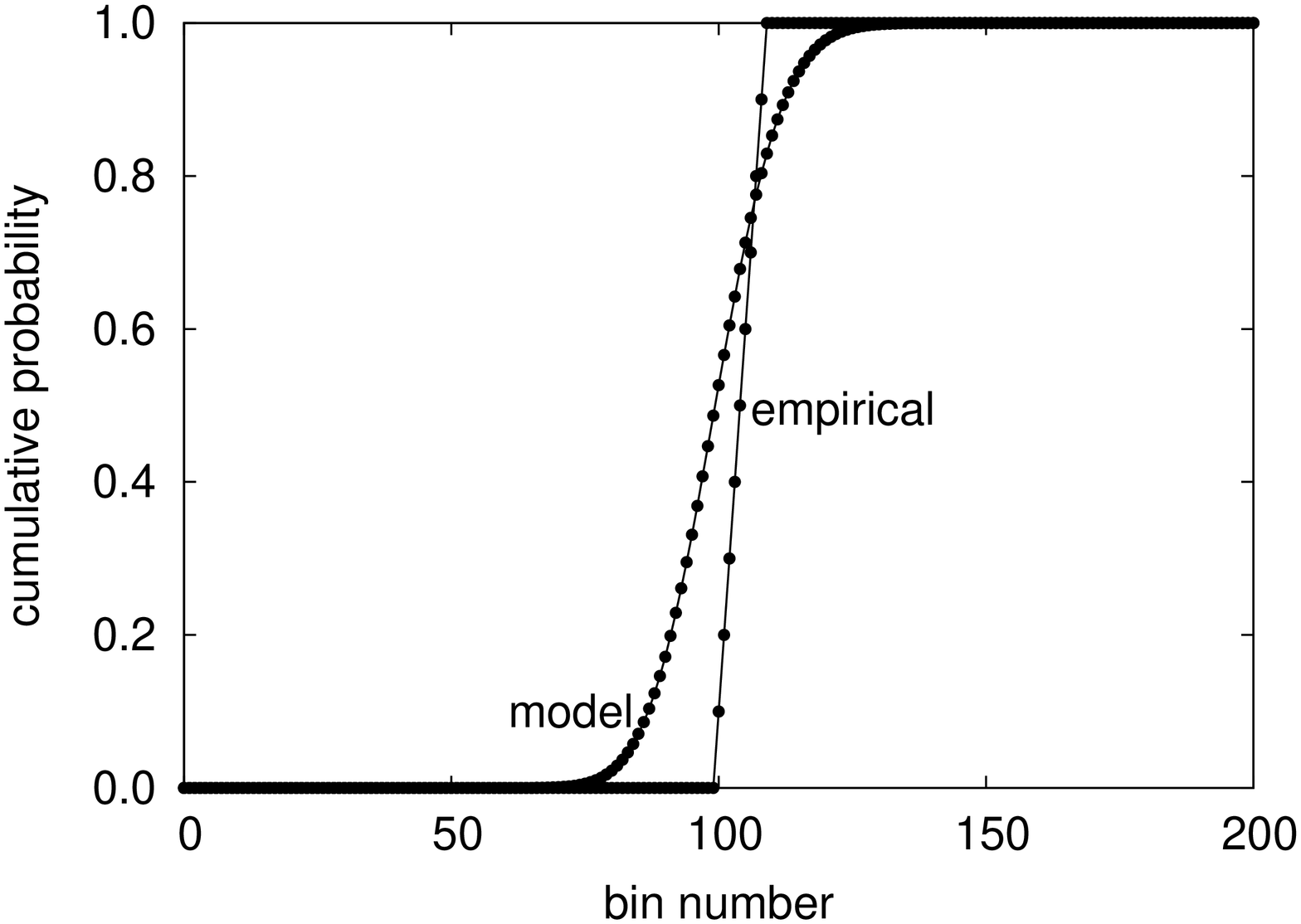}}}
\\\vspace{.1in}
\caption{Cumulative proportions associated with the bins for the observations}
\label{observedcmf}
\end{center}
\end{figure}

\begin{figure}
\begin{center}
\rotatebox{-90}{\scalebox{.47}{\includegraphics{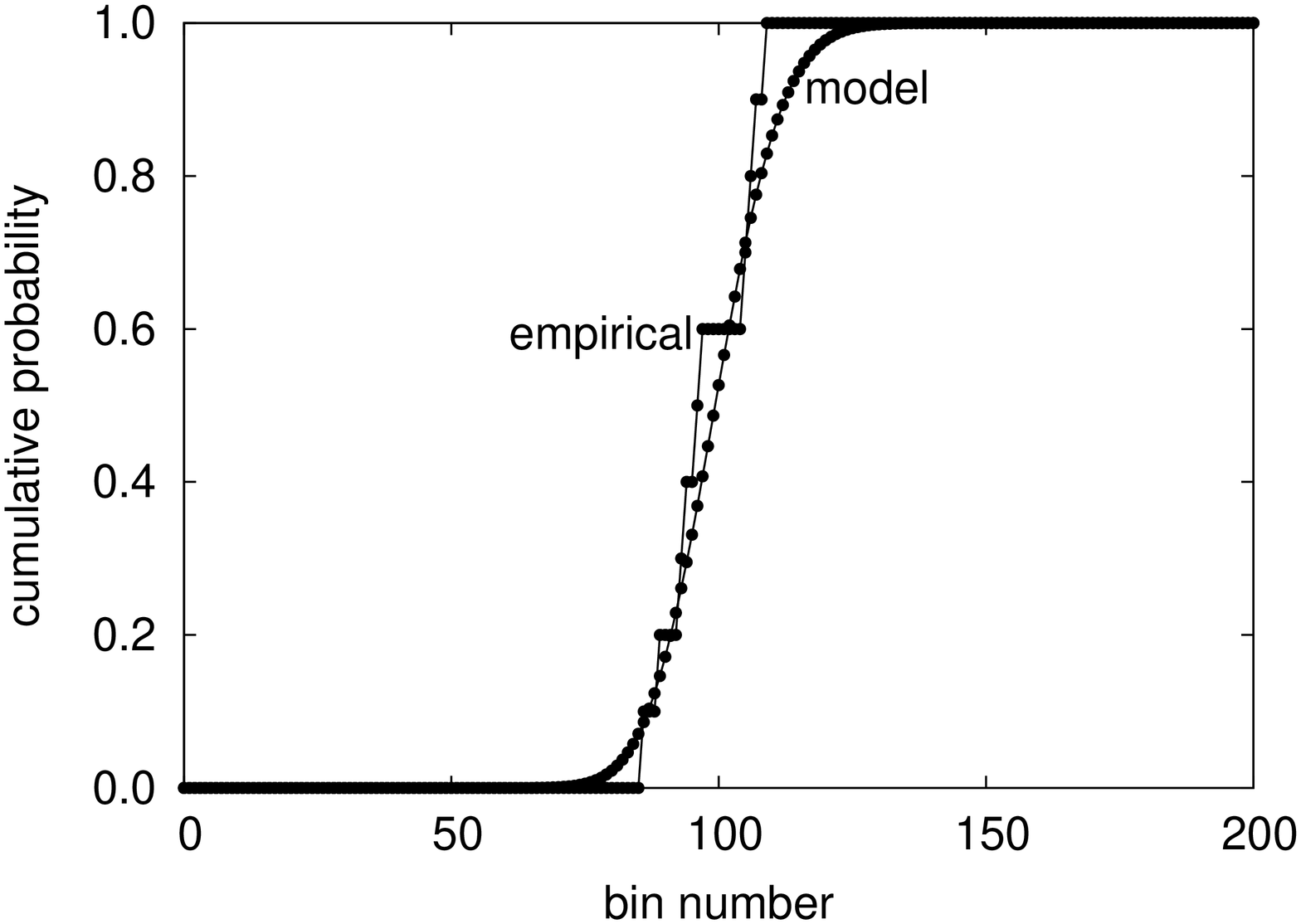}}}
\\\vspace{.1in}
\caption{Cumulative proportions associated with the bins for the simulation
         from Figure~\ref{simulatedpmf}}
\label{simulatedcmf}
\end{center}
\end{figure}

Figure~\ref{observedcmf} plots the cumulative model probabilities \
$p_0^{(0)}$,\; $p_0^{(0)}+p_0^{(1)}$,\; $p_0^{(0)}+p_0^{(1)}+p_0^{(2)}$,
\ \dots\ along with the cumulative observed proportions \
$\hat{p}^{(0)}$,\; $\hat{p}^{(0)}+\hat{p}^{(1)}$,\;
$\hat{p}^{(0)}+\hat{p}^{(1)}+\hat{p}^{(2)}$, \ \dots.
Figure~\ref{simulatedcmf} plots the cumulative model probabilities \
$p_0^{(0)}$,\; $p_0^{(0)}+p_0^{(1)}$,\; $p_0^{(0)}+p_0^{(1)}+p_0^{(2)}$,
\ \dots\ along with analogues of the cumulative proportions \
$\hat{p}^{(0)}$,\; $\hat{p}^{(0)}+\hat{p}^{(1)}$,\;
$\hat{p}^{(0)}+\hat{p}^{(1)}+\hat{p}^{(2)}$, \ \dots\
for the simulation generating 10 i.i.d.\ draws according to the model.

\subsection{A test of Hardy-Weinberg equilibrium}

In a population with suitably random mating,
the proportions of pairs of Rhesus haplotypes in members of the population
(each member has one pair) can be expected to follow the Hardy-Weinberg law
discussed by~\cite{guo-thompson},
namely to arise via random sampling from the model
\begin{equation}
\label{hw}
p_0^{(j,k)}(\theta_1, \theta_2, \dots, \theta_9)
= \left\{ \begin{array}{cl}
          2 \cdot \theta_j \cdot \theta_k, & j > k \\
          (\theta_k)^2, & j = k
  \end{array} \right.
\end{equation}
for $j,k = 1$,~$2$, \dots,~$9$ with $j \ge k$, under the constraint that
\begin{equation}
\sum_{j=1}^9 \theta_j = 1,
\end{equation}
where the parameters $\theta_1$,~$\theta_2$, \dots,~$\theta_9$
are the proportions of the nine Rhesus haplotypes in the population
(naturally, their maximum-likelihood estimates are the proportions
of the haplotypes in the given data).
For $j,k = 1$,~$2$, \dots,~$9$ with $j \ge k$, therefore,
$p_0^{(j,k)}$ is the expected probability that the pair of haplotypes
in the genome of an individual is the pair $j$ and $k$,
given the parameters $\theta_1$,~$\theta_2$, \dots,~$\theta_9$.

In this formulation, the hypothesis of suitably random mating entails that
the members of the sample population are i.i.d.\ draws from the model specified
in~(\ref{hw}); if a goodness-of-fit statistic rejects the model
with high confidence, then we can be confident that mating
has not been suitably random.

\begin{figure}
\begin{center}
\rotatebox{-90}{\scalebox{.47}{\includegraphics{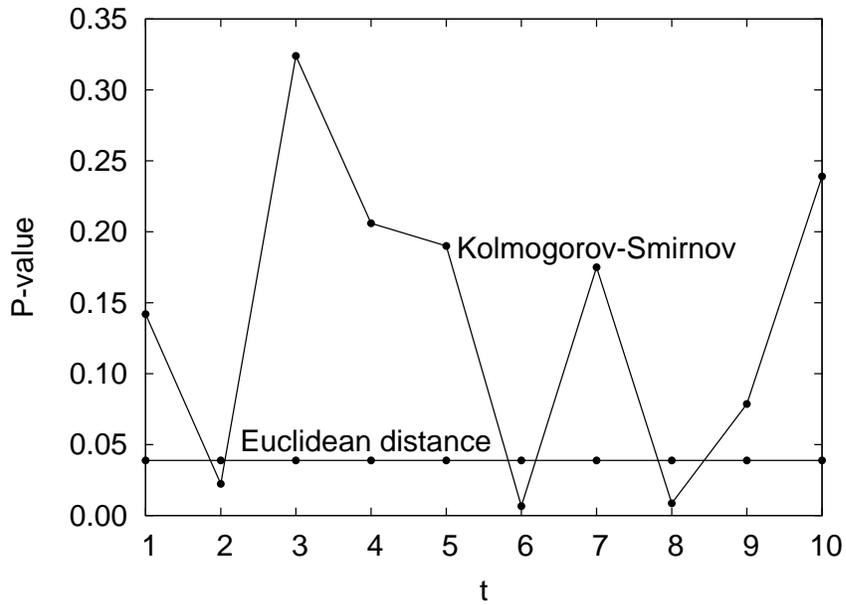}}}
\\\vspace{.1in}
\caption{P-values for Table~\ref{hwt} to be consistent with formula~(\ref{hw})}
\label{phwt}
\end{center}
\end{figure}

Table~\ref{hwt} provides data on $n = 8297$ individuals;
we duplicated Figure~3 of~\cite{guo-thompson} to obtain Table~\ref{hwt}.
Figure~\ref{phwt} plots the associated P-values,
each computed via 90,000 Monte-Carlo simulations.
The Kolmogorov-Smirnov statistic depends on the ordering of the bins;
for the first trial $t=1$ in Figure~\ref{phwt},
the order of the bins is the lexicographical ordering, namely
$(1,1)$,~$(2,1)$, $(2,2)$, $(3,1)$, $(3,2)$, $(3,3)$, \dots, $(9,9)$.
The nine trials $t = 2$,~$3$, \dots, $10$ displayed in Figure~\ref{phwt}
use pseudorandom orderings of the bins.
Please note that the Euclidean distance does not depend on the ordering.

Generally, a more powerful statistic produces lower P-values.
In Figure~\ref{phwt}, the P-values for the Kolmogorov-Smirnov statistic are
sometimes lower, sometimes higher than the P-values for the Euclidean distance.
There is no particularly natural ordering of the bins for~Figure~\ref{phwt};
Figure~\ref{phwt} displays 10 different orderings corresponding
to 10 different trials.
Figure~\ref{phwt} demonstrates that the Euclidean distance is more reliable
than the Kolmogorov-Smirnov statistic
when there is no natural ordering (or partial order) for the bins.

\begin{remark}
The P-values for classical goodness-of-fit statistics are substantially higher;
the classical statistics are less powerful for this example.
The P-values, each computed via 4,000,000 Monte-Carlo simulations, are
\begin{itemize}
\item Euclidean distance: .039
\item $\chi^2$: .693
\item $G^2$ (the log--likelihood-ratio): .600
\item Freeman-Tukey (the Hellinger distance): .562
\end{itemize}
For definitions and further discussion of the $\chi^2$, $G^2$,
and Freeman-Tukey statistics, see Section~4.5 of~\cite{perkins-tygert-ward3}.
Like the Euclidean distance, the $\chi^2$, $G^2$, and Freeman-Tukey statistics
are all invariant to the ordering of the bins.
\end{remark}

\begin{table}
\caption{Frequencies of pairs of Rhesus haplotypes}
\label{hwt}
\begin{center}
\hspace{3.5pc}$k$\\\vspace{4pt}
$j$
\begin{tabular}{c||c|c|c|c|c|c|c|c|c}
\hspace{-1pc}$_{j\hspace{-.3pc}}\diagdown{}^{\hspace{-.35pc}k}$\hspace{-1.3pc}
& 1 & 2 & 3 & 4 & 5 & 6 & 7 & 8 & 9 \\
\hline\hline
1 & 1236 &&&&&&& \\\hline
2 & 120 & 3 &&&&&&& \\\hline
3 & 18 & 0 & 0 &&&&& \\\hline
4 & 982 & 55 & 7 & 249 &&&& \\\hline
5 & 32 & 1 & 0 & 12 & 0 &&& \\\hline
6 & 2582 & 132 & 20 & 1162 & 29 & 1312 && \\\hline
7 & 6 & 0 & 0 & 4 & 0 & 4 & 0 & \\\hline
8 & 2 & 0 & 0 & 0 & 0 & 0 & 0 & 0 \\\hline
9 & 115 & 5 & 2 & 53 & 1 & 149 & 0 & 0 & 4
\end{tabular}
\end{center}
\end{table}

\subsection{A test of uniformity}

\begin{figure}
\begin{center}
\rotatebox{-90}{\scalebox{.47}{\includegraphics{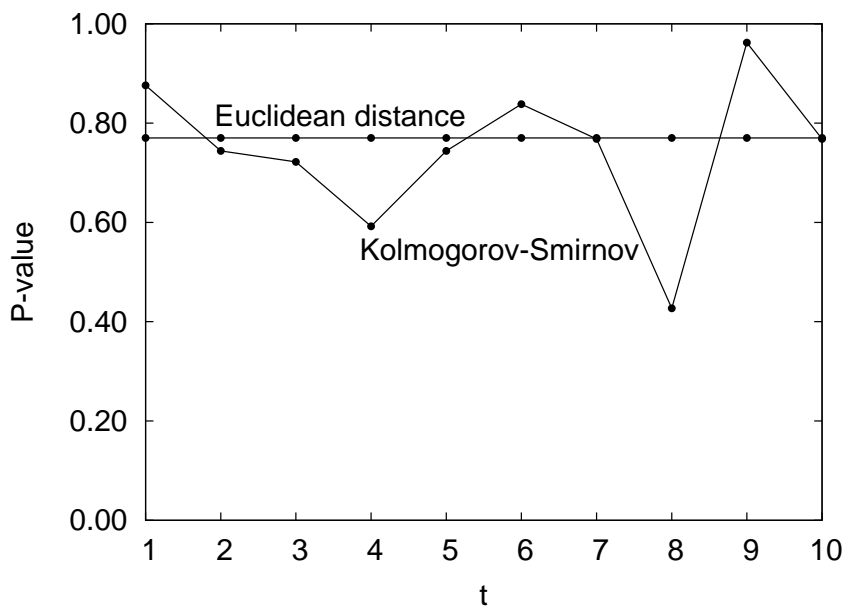}}}
\\\vspace{.1in}
\caption{P-values for Table~\ref{skittlest} to be consistent
         with the model displayed in Table~\ref{skittlestu}}
\label{pskittlest}
\end{center}
\end{figure}

Table~\ref{skittlest} duplicates Table~1 of~\cite{gilchrist},
giving the colors of the $n = 62$ pieces of candy in a 2.17 ounce bag.
Figure~\ref{pskittlest} plots the P-values for Table~\ref{skittlest} to be
consistent up to expected random fluctuations with Table~\ref{skittlestu},
the model of uniform proportions.
We computed each P-value via 4,000,000 Monte-Carlo simulations.
The Kolmogorov-Smirnov statistic depends on the ordering of the bins;
the ten trials $t = 1$,~$2$, \dots, $10$ displayed in Figure~\ref{pskittlest}
use pseudorandom orderings of the bins.
The Euclidean distance does not depend on the ordering.

Generally, a more powerful statistic produces lower P-values.
In Figure~\ref{pskittlest}, the P-values for the Kolmogorov-Smirnov statistic
are sometimes lower, sometimes higher than the P-values
for the Euclidean distance.
There is no particularly natural ordering of the bins
for Table~\ref{skittlestu};
Figure~\ref{pskittlest} displays 10 different pseudorandom orderings
corresponding to 10 different trials.
Figure~\ref{pskittlest} illustrates that the Euclidean distance
is more reliable than the Kolmogorov-Smirnov statistic
when there is no natural ordering (or partial order) for the bins.

\pagebreak

\begin{remark}
Table~\ref{skittlest} provides a possible means for ordering the bins.
However, such an ordering will depend on the observed data.
Using a data-dependent ordering can profoundly alter the nature
of the goodness-of-fit statistic; see Remark~\ref{l1}.
\end{remark}

\begin{remark}
Like the Euclidean distance, many classical goodness-of-fit statistics
are invariant to the ordering of the bins.
The following are P-values, each computed
via 4,000,000 Monte-Carlo simulations:
\begin{itemize}
\item Euclidean distance: .770
\item $\chi^2$: .770
\item $G^2$ (the log--likelihood-ratio): .766
\item Freeman-Tukey (the Hellinger distance): .755
\end{itemize}
For definitions and further discussion of the $\chi^2$, $G^2$,
and Freeman-Tukey statistics, see Section~2 of~\cite{perkins-tygert-ward3}.
For this example, the Euclidean distance and the $\chi^2$ statistic
produce exactly the same P-values:
for the model of homogeneous proportions, displayed in Table~\ref{skittlestu},
the Euclidean distance is directly proportional to the square root
of the $\chi^2$ statistic,
and hence the Euclidean distance is a strictly increasing function of $\chi^2$.
\end{remark}

\begin{table}
\caption{Observed frequencies of colors of candies in a 2.17 ounce bag}
\label{skittlest}
\begin{center}
\begin{tabular}{ccccccc}
 {\it color} && red & orange & yellow & green & violet \\
{\it number} &&  15 &      9 &     14 &    11 &     13
\end{tabular}
\end{center}
\end{table}

\begin{table}
\caption{Expected frequencies of colors of candies in a 2.17 ounce bag}
\label{skittlestu}
\begin{center}
\begin{tabular}{ccccccc}
 {\it color} &&  red & orange & yellow & green & violet \\
{\it number} && 12.4 &   12.4 &   12.4 &  12.4 &   12.4
\end{tabular}
\end{center}
\end{table}

\section*{Acknowledgements}
\addcontentsline{toc}{section}{\protect\numberline{}Acknowledgements}

We would like to thank Alex Barnett, G\'erard Ben Arous, James Berger,
Tony Cai, Sourav Chatterjee, Ronald Raphael Coifman, Ingrid Daubechies,
Jianqing Fan, Jiayang Gao, Andrew Gelman, Leslie Greengard, Peter W. Jones,
Deborah Mayo, Peter McCullagh, Michael O'Neil, Ron Peled, William Perkins,
William H. Press, Vladimir Rokhlin, Joseph Romano, Gary Simon, Amit Singer,
Michael Stein, Stephen Stigler, Joel Tropp, Larry Wasserman,
and Douglas A. Wolfe.
This work was supported in part by Alfred P. Sloan Research Fellowships,
a Donald D. Harrington Faculty Fellowship,
and a DARPA Young Faculty Award.


\addcontentsline{toc}{section}{\protect\numberline{}References}
\bibliographystyle{asamod.bst}
\bibliography{stat}

\begin{thebibliography}{22}
\newcommand{\enquote}[1]{``#1''}
\expandafter\ifx\csname natexlab\endcsname\relax\def\natexlab#1{#1}\fi

\bibitem[{Ampadu(2008)}]{ampadu}
Ampadu, C. (2008), On the powers of some new chi-square type statistics,
  \textit{Far East J. Theoretical Statist.}, \textbf{26}, 59--72.

\bibitem[{Ampadu et~al.(2009)Ampadu, Wang, and Steele}]{ampadu-wang-steele}
Ampadu, C., Wang, D., and Steele, M. (2009), Simulated power of some discrete
  goodness-of-fit test statistics for testing the null hypothesis of a zig-zag
  distribution, \textit{Far East J. Theoretical Statist.}, \textbf{28},
  157--171.

\bibitem[{Best and Rayner(1997)}]{best-rayner}
Best, D.~J. and Rayner, J. C.~W. (1997), Goodness-of-fit for the ordered
  categories discrete uniform distribution, \textit{Comm. Statist. Theory
  Meth.}, \textbf{26}, 899--909.

\bibitem[{Choulakian et~al.(1994)Choulakian, Lockhart, and
  Stephens}]{choulakian-lockhart-stephens}
Choulakian, V., Lockhart, R.~A., and Stephens, M.~A. (1994),
  Cram\'er--von-Mises statistics for discrete distributions, \textit{Canadian
  J. Statist.}, \textbf{22}, 125--137.

\bibitem[{D'Agostino and Stephens(1986)}]{dagostino-stephens}
D'Agostino, R.~B. and Stephens, M.~A. (1986), \textit{Goodness-of-Fit
  Techniques}, New York: Marcel Dekker.

\bibitem[{Durbin(1972)}]{durbin}
Durbin, J. (1972), \textit{Distribution Theory for Tests Based on the Sample
  Distribution Function}, CBMS-NSF Regional Conference Series in Applied
  Mathematics, Philadelphia: Society for Industrial and Applied Mathematics.

\bibitem[{From(1996)}]{from}
From, S.~G. (1996), A new goodness-of-fit test for the equality of multinomial
  cell probabilities versus trend alternatives, \textit{Comm. Statist. Theory
  Meth.}, \textbf{25}, 3167--3183.

\bibitem[{Gilchrist(2010)}]{gilchrist}
Gilchrist, E. (2010), A sweet approach to teaching the one-variable chi-square
  test, \textit{Communication Teacher}, \textbf{24}, 14--18.

\bibitem[{Guo and Thompson(1992)}]{guo-thompson}
Guo, S.~W. and Thompson, E.~A. (1992), Performing the exact test of
  {H}ardy-{W}einberg proportion for multiple alleles, \textit{Biometrics},
  \textbf{48}, 361--372.

\bibitem[{Haschenburger and Spinelli(2005)}]{haschenburger-spinelli}
Haschenburger, J.~K. and Spinelli, J.~J. (2005), Assessing the goodness-of-fit
  of statistical distributions when data are grouped, \textit{Math. Geology},
  \textbf{37}, 261--276.

\bibitem[{Hoeffding(1965)}]{hoeffding}
Hoeffding, W. (1965), Asymptotically optimal tests for multinomial
  distributions, \textit{Ann. Math. Statist.}, \textbf{36}, 369--401.

\bibitem[{Horn(1977)}]{horn}
Horn, S.~D. (1977), Goodness-of-fit tests for discrete data: a review and an
  application to a health impairment scale, \textit{Biometrics}, \textbf{33},
  237--247.

\bibitem[{Lockhart et~al.(2007)Lockhart, Spinelli, and
  Stephens}]{lockhart-spinelli-stephens}
Lockhart, R.~A., Spinelli, J.~J., and Stephens, M.~A. (2007),
  Cram\'er--von-Mises statistics for discrete distributions with unknown
  parameters, \textit{Canadian J. Statist.}, \textbf{35}, 125--133.

\bibitem[{Marsaglia et~al.(2003)Marsaglia, Tsang, and
  Wang}]{marsaglia-tsang-wang}
Marsaglia, G., Tsang, W.~W., and Wang, J. (2003), Evaluating {K}olmogorov's
  distribution, \textit{J. Statist. Soft.}, \textbf{8}, 1--4.

\bibitem[{Perkins et~al.(2011{\natexlab{a}})Perkins, Tygert, and
  Ward}]{perkins-tygert-ward3}
Perkins, W., Tygert, M., and Ward, R. (2011{\natexlab{a}}), $\chi^2$ and
  classical exact tests often wildly misreport significance; the remedy lies in
  computers, Tech. Rep. 1108.4126, arXiv,
  http://cims.nyu.edu/$\sim$tygert/abbreviated.pdf.

\bibitem[{Perkins et~al.(2011{\natexlab{b}})Perkins, Tygert, and
  Ward}]{perkins-tygert-ward2}
Perkins, W., Tygert, M., and Ward, R. (2011{\natexlab{b}}), Computing the
  confidence levels for a root-mean-square test of goodness-of-fit, {II}, Tech.
  Rep. 1009.2260, arXiv.

\bibitem[{Pettitt and Stephens(1977)}]{pettitt-stephens}
Pettitt, A.~N. and Stephens, M.~A. (1977), The {K}olmogorov-{S}mirnov
  goodness-of-fit statistic with discrete and grouped data,
  \textit{Technometrics}, \textbf{19}, 205--210.

\bibitem[{Press et~al.(2007)Press, Teukolsky, Vetterling, and
  Flannery}]{press-teukolsky-vetterling-flannery}
Press, W., Teukolsky, S., Vetterling, W., and Flannery, B. (2007),
  \textit{Numerical Recipes}, Cambridge, UK: Cambridge University Press, 3rd
  ed.

\bibitem[{Rao(2002)}]{rao}
Rao, C.~R. (2002), {K}arl {P}earson chi-square test: {T}he dawn of statistical
  inference, In \textit{Goodness-of-Fit Tests and Model Validity} (eds
  Huber-Carol, C., Balakrishnan, N., Nikulin, M.~S., and Mesbah, M.), pp.
  9--24, Boston: Birkh\"auser.

\bibitem[{R\'enyi(1953)}]{renyi}
R\'enyi, A. (1953), On the theory of order statistics, \textit{Acta Math. Acad.
  Sci. Hungar.}, \textbf{4}, 191--231.

\bibitem[{Steele and Chaseling(2006)}]{steele-chaseling}
Steele, M. and Chaseling, J. (2006), Powers of discrete goodness-of-fit
  statistics for a uniform null against a selection of alternative
  distributions, \textit{Comm. Statist. Simul. Comput.}, \textbf{35},
  1067--1075.

\bibitem[{Stephens(1970)}]{stephens2}
Stephens, M.~A. (1970), Use of the {K}olmogorov-{S}mirnov,
  {C}ram\'er--{V}on-{M}ises and related statistics without extensive tables,
  \textit{J. Roy. Statist. Soc. Ser. B}, \textbf{32}, 115--122.

\end{thebibliography}

\end{document}